# Comparison of mean-field based theoretical analysis methods for SIS model

Jiaquan Zhang[1], Dan Lu[1], Shunkun Yang[1]*

[1]*School of Reliability and Systems Engineering, Beihang University, Beijing, China*

*ysk@buaa.edu.cn

## Abstract

Epidemic spreading has been intensively studied in SIS epidemic model. Although the mean-field theory of SIS model has been widely used in the research, there is a lack of comparative results between different theoretical calculations, and the differences between them should be systematically explained. In this paper, we have compared different theoretical solutions for mean-field theory and explained the underlying reason. We first describe the differences between different equations for mean-field theory in different networks. The results show that the difference between mean-field reaction equations is due to the different probability consideration for the infection process. This finding will help us to design better theoretical solutions for epidemic models.

# Introduction

In research of epidemic spreading, SIS (Susceptible-Infected-Susceptible) model [1,2] is widely applied because of its balance between generality and simplicity. Through large-scale simulations and theoretical calculations [3-12], there have been many findings in epidemic propagation based on SIS model, including the infection threshold $\lambda_c$ [7,13-15] on networks. Classical SIS model can also be extended to describe the real diseases [16,17], to predict the infection density of disease propagation and to reduce the duration of epidemic outbreak (e.g., SARS, AIDS) [18,19].

Among these studies, mean-field theory is one of the popular analysis methods for SIS epidemic models. The classical mean-field approximation [20] is mainly applied to homogeneous networks, and there exists a finite critical value $\lambda_c$, which is the inverse of the average degree $\langle k \rangle$ [21]. Pastor-Satorras and Vespignani proposed the heterogeneous mean-field (HMF) theory [3], which is then applied to solve the epidemic threshold on uncorrelated networks (i.e., $\lambda_c = \langle k \rangle / \langle k^2 \rangle$) [5, 6]. The finite epidemic threshold $\lambda_c > 0$ for SF networks can also be computed based on the HMF theory by introducing a saturation function [22].

The mean-field theory can solve the epidemic threshold on networks, with the considerations of complex conditions. Through analyzing the immunization strategies of SIS model with considering the degree information [23], the epidemic threshold on degree-correlated networks is found to be lower than that on degree-uncorrelated networks [24]. In addition, the HMF theory is used to consider the SIS dynamics and contact process on annealed directed scale-free networks with degree-degree correlation [25], where the threshold $\lambda_c = 0$ for $\gamma \leq 3$ as that in the undirected networks.

The mean-field theory can also be used to analyze the dynamics on spatial networks. Meanwhile, the quenched mean-field theory (QMF) is specially proposed to calculate the corresponding epidemic threshold on quenched networks. In the case of quenched undirected networks, the epidemic threshold does not exist on networks with infinite size [26]. The quenched mean-field approximation [28] with spectral decomposition can be used to understand the rare-region effects of SIS model on weighted SF networks by considering the variations of the quenched network topology [30]. It shows that $\lambda_c$ vanishes within the limitation of $N \to \infty$ [29]. Considering the irreversible fluctuation, which was ignored in the QMF theory, the epidemic threshold is nonzero in the un-clustered SF networks [27] as opposed to the QMF theory.

Given the wide application of theoretical analysis during SIS epidemic propagation, there is lack of comparison between different mean-field theory calculations. Specifically, most studies focus on the critical condition of epidemics breakout, while less attention is focused on the study of infection density, on which different theoretical solutions show distinct results. In addition, the differences between them have not been systematically analyzed.

Through the comparison between theoretical calculations, we find some differences on random-regular networks. The root cause which could explain the differences is that item $\beta \langle k \rangle \rho(t)$ describing the infection process in classical mean-field equation has different probability consideration from that in simulations. Therefore, the improved mean-field equation is proposed based on the analysis of the root cause. In addition, the results of the comparison between the simulations and theoretical verification demonstrate that the modified equation is more accordant with simulations than the unimproved one on different kinds of networks [31,32].

The contents of this paper are divided into the following sections. In section 2, we propose the modified mean-field equation by considering the differences between the classical MF equation and simulations. Section 3 is devoted to analyze the cause of the difference between classical MF equation and improved MF equation. Section 4 demonstrates the advantage of improved MF equation through detailed comparison between theoretical calculations of the two equations. In section 5, we study a test case on small-world network to show the generality of the improved MF equation.

## Classical and modified MF equation of SIS model

The most widely accepted mean-field reaction equation of SIS model is classical MF equation (equation 1). It ignores the differences between nodes and it is suited for homogeneous networks. However, we found that this equation is different from simulations when we compared the steady-state infection density $\rho(t)$. According to our analysis, the reason is not related to inhomogeneity or sparsity of the network, but is the different probability considerations for the infection process. As a result, we modified the corresponding parts of the classical MF equation and rewrote the equation (equation 2), which considers that a node being infected by its neighbors is not independent.

$$\frac{\partial \rho(t)}{\partial t} = \beta(t)\langle k \rangle \rho(t)(1-\rho(t)) - \delta(t)\rho(t). \tag{1}$$

$$\frac{\partial \rho(t)}{\partial t} = (1-\rho(t))(1-(1-\beta)^{\langle k \rangle \rho(t)}) - \delta(t)\rho(t). \tag{2}$$

Where, $\rho(t)$ is the infection density. $\langle k \rangle$ is the average degree of the network. The infection rate $\lambda(t) = \beta(t)/\delta(t)$.

## Analysis of the difference between two different equations

In this section, we first give detailed comparison of the difference between equation (1) and simulations of SIS model on both ER and RR networks. Then we analyze the cause of the difference and showed theoretical analysis of the difference between the two equations. Usually there exists some differences between simulations and theoretical calculation of equation (1) on ER networks. The most obvious difference is the infection density $\rho(t)$ in steady state. To conveniently show the difference, we performed simulations and theoretical calculation of equation with the assumption that $\beta(t)$ and $\delta(t)$ are both constant, which means they do not change with time.

For theoretical solution (blue line in Fig. 1), the stationary condition of equation (1)

means $\frac{\partial \rho(t)}{\partial t}=0$. Then the stationary $\rho(t)$ is $1-\frac{\delta(t)}{\beta(t)\langle k\rangle}$, which is $1-\frac{\delta}{\beta\langle k\rangle}$ in this case. Noticing that if $\lambda=\frac{\beta}{\delta}$ is a constant value, stationary $\rho(t)$ can also remain constant.

For simulations, at each step, we first searched the neighbors for each susceptible node. If we find an infectious neighbor, then a random number $p\in(0,1)$ is created; if $p<\beta$, this susceptible node would be infected next step and this local searching loop stopped; if $p\geq\beta$, we continue searching until every neighbor has been handled.

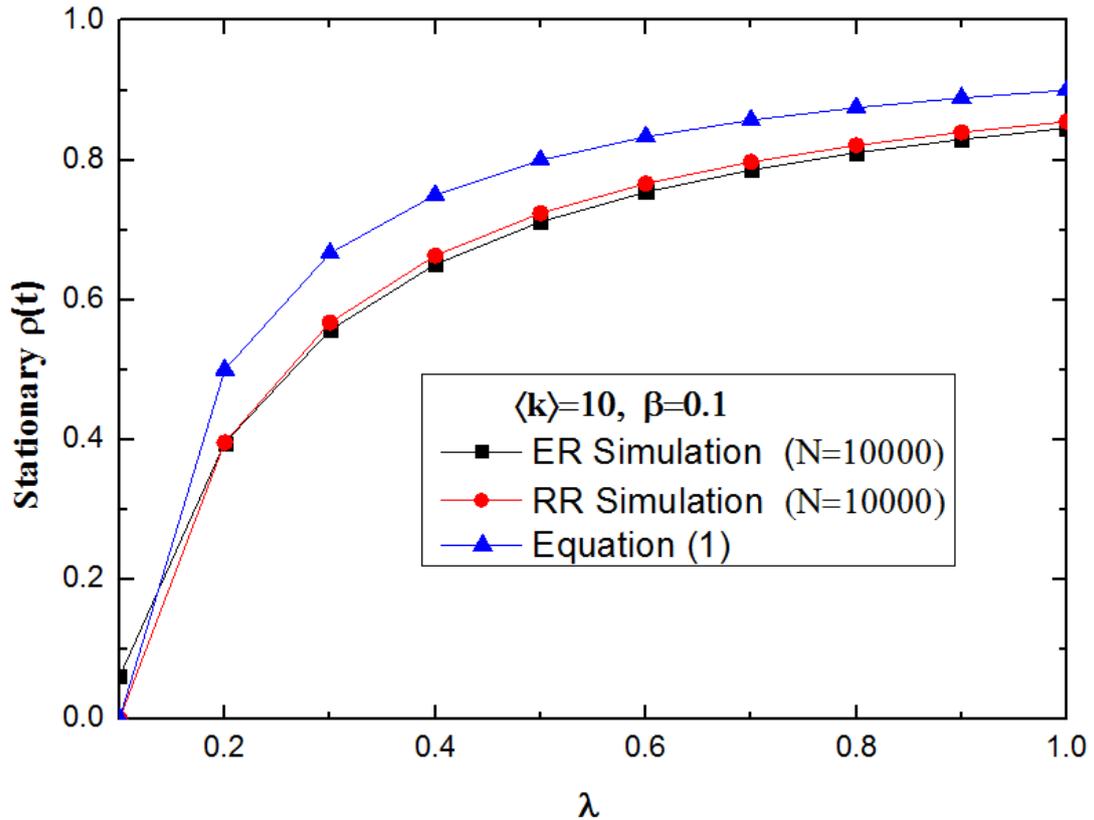

**Figure 1 Stationary $\rho(t)$ of equation 1, and simulations on ER and RR networks with N=10000 and $\langle k\rangle$ =10.** Simulation results are averages over 100 realizations. $\beta=0.1$, infection rate $\lambda$=0.1, 0.2, ..., 1.0.

As shown in Fig. 1, firstly the simulation results on ER networks are obviously different from theoretical results of equation (1), and the differences decrease with the increasing of $\lambda$. Considering that ER network is not strictly homogeneous, which means all nodes do not have the same number of links, we may consider that the difference is caused by the inhomogeneity of ER network.

Theoretically, equation (1) should perform well on random-regular network because it is strictly homogeneous, which means the RR network meets all the assumptions and conditions that equation (1) requires. Each node in RR network has the same number of links, hence there should be no difference between simulation and equation (1). Actually, theoretical solution by using the average degree $\langle k\rangle$ is a solution for random-regular network rather than ER network. For comparison, we carried out same simulations on RR networks with N=10000 and $\langle k \rangle$=10 as well. Also shown in Fig 1, obviously there still exists differences between simulations and theoretical results even when we simulated the SIS epidemic propagation on dense random-regular networks. Although the simulation results of RR network are closer to theoretical solutions of equation (1) than those of ER network, they are still obviously different from the simulations. Moreover, the differences between simulation results on ER networks and RR networks remain almost constant.

The analysis and observations above demonstrate that this difference is not caused by inhomogeneity or sparsity of networks. The root cause actually is equation (1) and simulation have different probability considerations for the infection process. Equation (1) shows that the number of infectious nodes of the whole network at each step increases by $\beta(t)\langle k \rangle \rho(t)(1-\rho(t))$ and decreases by $\delta(t)\rho(t)$. Specifically, $(1-\rho(t))$ is the percentage of susceptible nodes; $\langle k \rangle \rho(t)$ is the number of infectious neighbors of each susceptible node; $\beta(t)$ is the probability that a node would be infected by an infectious neighbor. Each part of the formula has a reasonable meaning. However, the problem comes from $\beta(t)\langle k \rangle \rho(t)$, which indicates the probability of a susceptible node being infected by its several infectious neighbors. We found this is not consistent with the simulation. Because a node being infected by each of its infectious neighbors is not independent, simply adding the infectious rate $\beta(t)$ to calculate the total probability that a node being infected by its infectious neighbors is not correct. To elaborately demonstrate the problem above, the following is a simple case study.

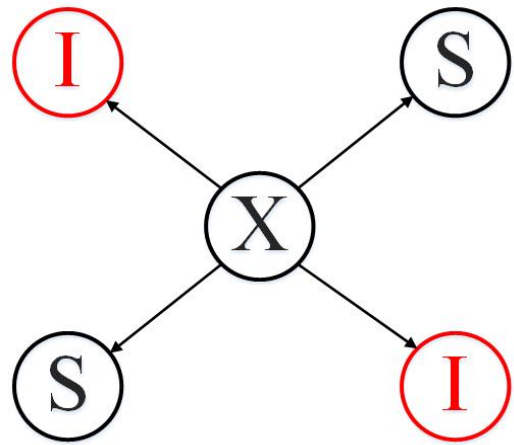

**Figure 2  A case to demonstrate the probability that a node being infected by its neighbors.** Node X is a susceptible node which has 4 neighbors, two of its neighbors are infected and the other two are susceptible.

For the case in Fig. 2, we assume that $\beta = 0.1$. Then according to equation (1), the probability that node X becoming infected in the next step is:
$$\beta \times \langle k \rangle \times \rho(t) = 2 \times 0.1 = 0.2 \quad (\langle k \rangle \times \rho(t) = 2 \text{ in this case}),$$
However, according to the simulation mentioned above, the probability that node X becoming infected is:
$$1-(1-\beta)^{\langle k \rangle \rho(t)} = 0.19.$$
This is the cause of why equation (1) is different from simulation. We therefore modified the classical MF equation of SIS model by changing $\beta \times \langle k \rangle \times \rho(t)$ to

$1-(1-\beta(t))^{\langle k\rangle \rho(t)}$. Then the improved mean-field reaction equation of SIS model should be rewritten as equation (2), and it is theoretically more consistent with simulation than equation (1). The following parts will demonstrate why equation (2) is better than equation (1) and discuss more about the differences between the two equations in detail.

## Comparison of theoretical calculations between two equations

We demonstrate the advantage of equation (2) through comparison from two aspects. Firstly, we compare the results from a reverse method by only considering the stationary condition, which is quite easy and convictive. Stationary condition means that $\frac{\partial \rho(t)}{\partial t}=0$, then equation (1) and equation (2) can be written as follows:

$$\beta(t)\langle k\rangle \rho(t)(1-\rho(t))-\delta(t)\rho(t)=0 \qquad (3)$$

$$(1-\rho(t))(1-(1-\beta)^{\langle k\rangle \rho(t)})-\delta(t)\rho(t)=0 \qquad (4)$$

Because we have got the simulation results of steady-state infection density $\rho(t)$, if the simulation results could give a true statement of equation (3) and (4), then it could demonstrate that which one is more consistent with the facts. According to Fig 1, for RR network with $\langle k\rangle = 10$ and N = 10000, when $\lambda=0.3$ ($\beta=0.1$) the $\rho(t)$ on stationary condition of simulation is 0.56747. Then the numerical results of the equation (3) and equation (4) is: $0.24545-0.18916=0.05629$, $0.19465-0.18916=0.00549$. Obviously, this can demonstrate that the equation (2) can better describe the process of epidemic. For all simulations, the results of stationary $\rho(t)$ could give true statements of equation (4) within very small differences. To avoid repetition, here only presents one of them.

Besides, the numerical comparison for stationary condition based on simulation results, we also calculated the whole process of equation (1) and (2) theoretically. Rather than finding a general solution of the two equations, we used MATLAB to calculate $\rho(t)$ under specific parameters (Fig. 3). To find out more about how differences between the two equations change with $\beta$ and $\delta$, we first compared several groups of results under different $\beta$ and $\delta$. Noticing that $\lambda=\frac{\beta}{\delta}$ remains the same value among these groups, this can eliminates the interference of $\lambda$.

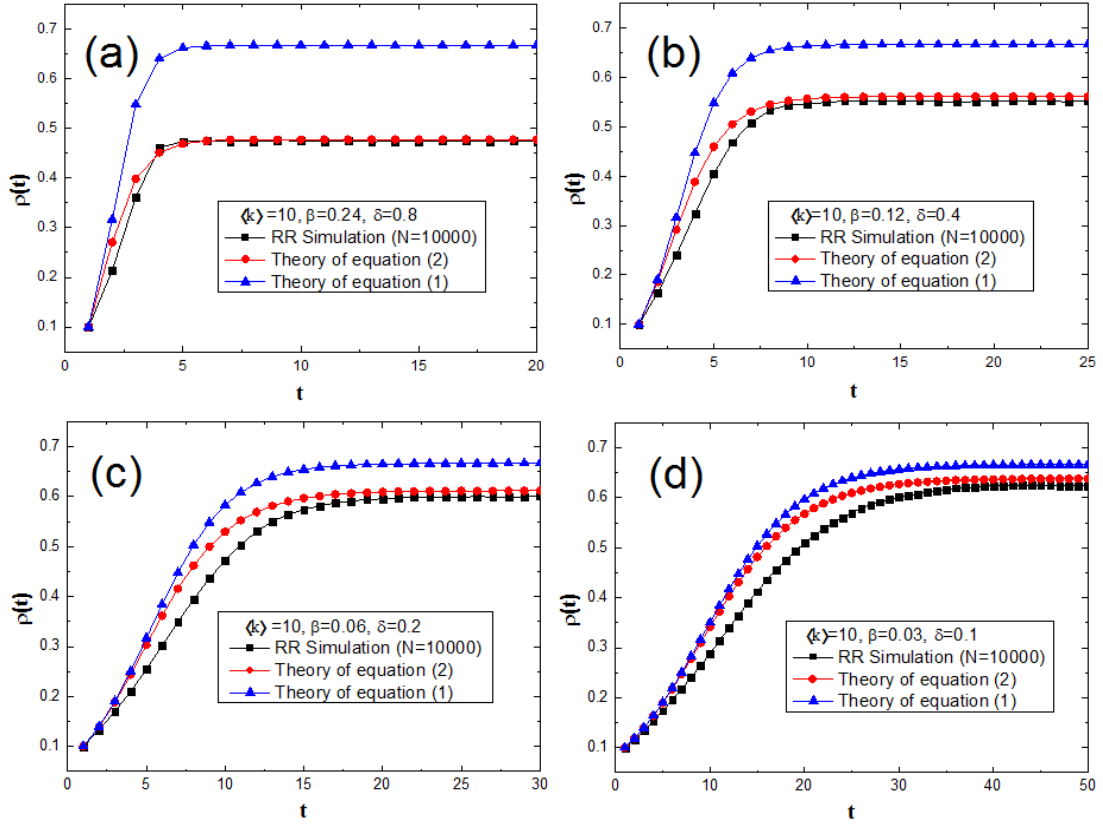

**Figure 3 Comparison between theoretical calculations of equation (1) and equation (2), $\lambda$ remains 0.3.** Simulation results are averaged over 100 realizations. **(a)** $\beta = 0.24$, $\delta = 0.8$. **(b)** $\beta = 0.12$, $\delta = 0.4$. **(c)** $\beta = 0.06$, $\delta = 0.2$. **(d)** $\beta = 0.03$, $\delta = 0.1$. Simulation results are averaged over 100 realizations

Observed from Fig. 3, the differences between simulations on RR networks and equation (2) are always far less than that of equation (1), especially in stationary condition. And for a certain $\lambda$, the theoretical results of equation (2) are more consistent with those of simulations when $\beta$ increases. Meanwhile, the differences between equation (2) and equation (1) decrease obviously with the increasing of $\beta$. To explain the relation between the two equations, we need to focus on the probability of each susceptible node being infected by its neighbors. Because this part is the most important difference between the two equations. As $\langle k \rangle \times \rho(t)$ indicates the number of infectious neighbors for each node, here we suppose it as $\alpha$. Then Taylor's expansion of $1-(1-\beta)^{\langle k \rangle \rho(t)}$ is:

$$1-(1-\beta)^{\alpha} = 1-[1-\alpha\beta+\frac{\alpha(\alpha-1)\beta^2}{2!}-\frac{\alpha(\alpha-1)(\alpha-2)\beta^3}{3!}+\cdots]$$

$$=\alpha\beta+\frac{\alpha(\alpha-1)\beta^2}{2!}-\frac{\alpha(\alpha-1)(\alpha-2)\beta^3}{3!}+\cdots$$

When $\beta \to 0$, the higher order items could be ignored and $1-(1-\beta)^{\alpha} \to \alpha\beta$, which means $\beta \times \langle k \rangle \times \rho(t) = 1-(1-\beta)^{\langle k \rangle \rho(t)}$. The two equations are now equal, hence the stationary $\rho(t)$ of the simulation is quite close to equation (1). Actually, the theoretical difference between the two equations is:

$$[1-(1-\beta)^\alpha]-\beta\alpha = \frac{\alpha(\alpha-1)\beta^2}{2!} - \frac{\alpha(\alpha-1)(\alpha-2)\beta^3}{3!} + \cdots.$$

For a given $\alpha$, the differences increase with the increasing of $\beta$, which we have addressed above in Fig. 3.

Fig. 4 is comparison of stationary $\rho(t)$ between the two equations under different $\lambda$. To avoid repeating information in Fig. 1, here we use $\langle k \rangle$=6 and $\langle k \rangle$=4 instead of $\langle k \rangle$=10. To demonstrate the advantage of equation (2), we took simulations on ER networks into comparison as well. The results show that both equation (1) and equation (2) are not well consistent with simulations on ER network with $\langle k \rangle$=4, which is actually duo to the sparsity of the network. Although, equation (2) is still better than equation (1). For $\langle k \rangle$=6, it is more obvious that equation (2) performs better. Moreover, the differences between equation (2) and simulations remain almost constant with the increasing of $\lambda$.

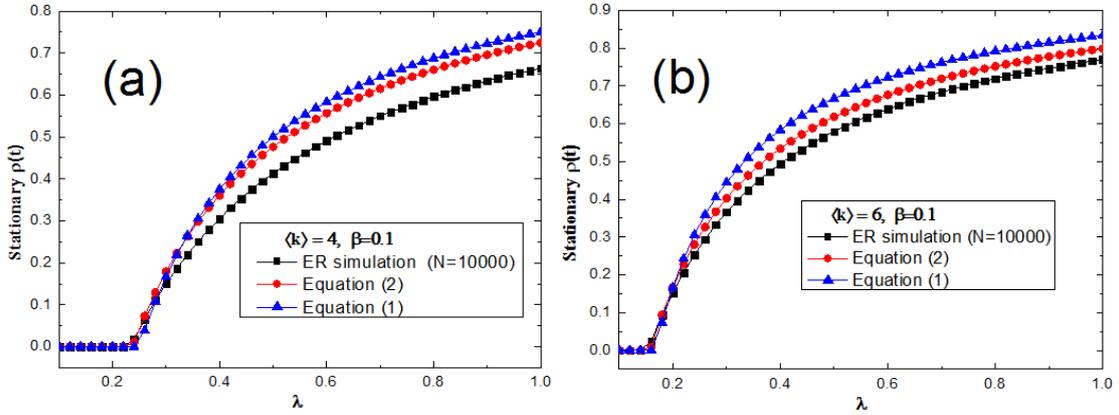

**Figure 4 Comparison between theoretical calculations of equation (1) and equation (2) on RR networks, to demonstrate which one is more consistent with simulation. (a)** N=10000, $\langle k \rangle$=4, $\beta$=0.1. **(b)** N=10000, $\langle k \rangle$=6, $\beta$=0.1.

Meanwhile, we compared the results of $\lambda_c$. As we know, $\lambda_c$ derived from equation (1) is $\frac{1}{\langle k \rangle}$ [2]. $\lambda_c$ is computed under the condition $\rho(t) \rightarrow 0$. When $\rho(t) \rightarrow 0$, $\rho(t) \times \langle k \rangle \rightarrow 0$, which means $1-(1-\beta(t))^{\langle k \rangle \rho(t)} = \beta \times \langle k \rangle \times \rho(t)$. Under such condition, equation (2) is equal to equation (1). As a result, equation (2) does not change the epidemic threshold $\lambda_c$ defined by equation (1), which can also observed from Figure 4.

## Test case on small-world networks

The above comparisons between two equations are based on RR networks, which seldom exist in real world. As a result, we show a test case on small-world networks with different reconnect rates (p). Fig. 5 demonstrates that equation (2) is better than equation (1) for the consistency with SIS models even on small-world networks. Moreover, the differences between equations (2) and simulations decrease with the increasing of $p$.

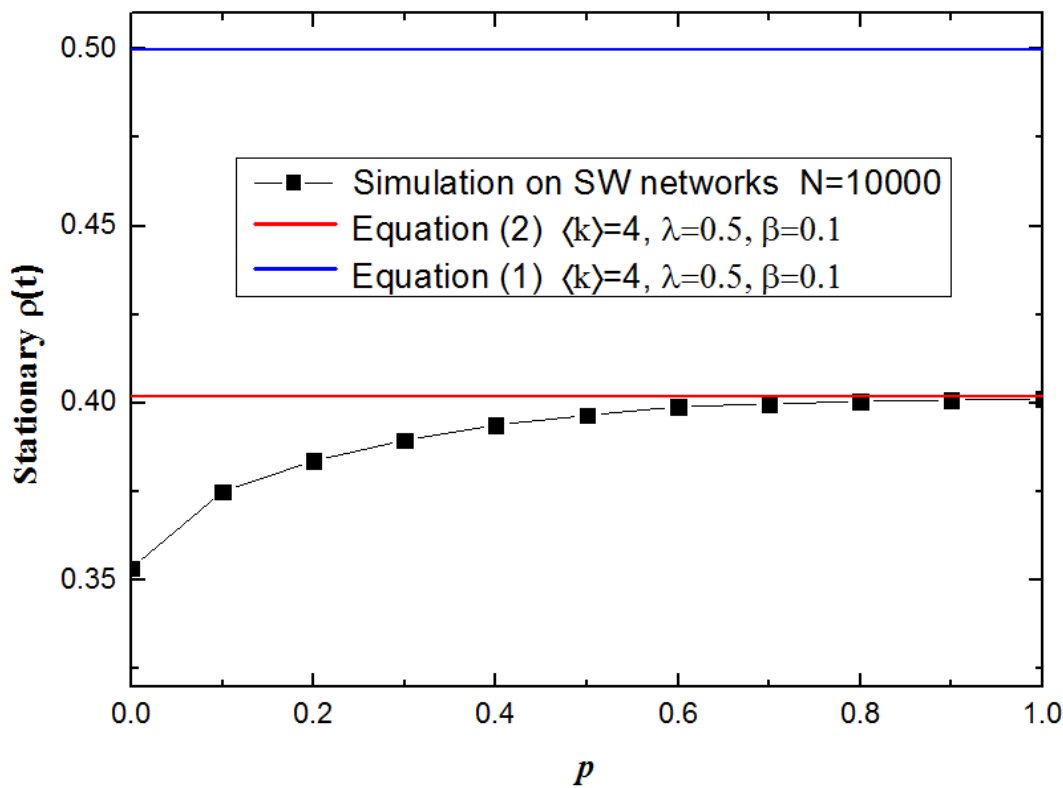

**Figure 5 Comparison between theoretical calculations of equation (1) and equation (2) on small-world networks with different p.** Simulation results are averaged over 100 realizations. $\lambda=0.5$, $\beta=0.1$, $N=10000$, $\langle k \rangle = 4$.

## Discussion and Conclusion

Equation (1) has another problem. As $\beta \times \langle k \rangle \times \rho(t)$ is the probability that a susceptible node being infected by its neighbors, it should not bigger than 1. However, it will over 1.0 when $\beta = 0.24$, $\langle k \rangle = 10$, $\rho(t) = 0.5$. If so, it not only causes differences but also cause mistakes due to the logical defect. While for $1-(1-\beta)^{\langle k \rangle \rho(t)}$ of equation 2, it can always represent the right meaning of the probability.

In this paper, we found that for SIS model, the difference between classical mean-field reaction equation and simulation is not caused by inhomogeneity or sparsity of network. By analyzing theoretical solutions and simulations in detail, we found that the root cause of the difference is due to probability consideration for the infection process. Then we showed a modified equation, which is more consistent with simulation. By comparing theoretical calculations of the two equations, we demonstrate that the improved equation preforms better than the classical one on different kinds of networks, like RR, ER and SW network. However, the two equations are equal in limiting conditions, which means the new equation does not change the epidemic threshold $\lambda_c$ defined by the classical equation.

# Acknowledgement

We thank the support from National Natural Science Foundation of China (Grant Nos. 61672080) and National Aerospace Science Foundation of China (Grant Nos. 2011ZD51055 and 2016ZD51031).